\documentclass{article}
\usepackage{amssymb}

\usepackage{graphicx}
\usepackage{amsmath}

\newtheorem{theorem}{Theorem}

\newtheorem{definition}[theorem]{Definition}
\newtheorem{example}[theorem]{Example}

\newtheorem{lemma}[theorem]{Lemma}

\newtheorem{proposition}[theorem]{Proposition}
\newtheorem{remark}[theorem]{Remark}

\newenvironment{proof}[1][Proof]{\textbf{#1.} }{\ \rule{0.5em}{0.5em}}
\input{tcilatex}

\begin{document}

\title{\textbf{Equivariant characteristic forms}\\
\textbf{on the bundle of connections}}
\date{}
\author{\textsc{R. Ferreiro P\'{e}rez} \\
Insituto de F\'{i}sica Aplicada, CSIC\\
C/ Serrano 144, 28006-Madrid\\
\emph{E-mail:} \texttt{roberto@iec.csic.es}}
\maketitle

\begin{abstract}
\noindent%
%
The characteristic forms on the bundle of connections of a principal bundle $%
P\rightarrow M$ of degree equal to or less than $\dim M$, determine the
characteristic classes of $P$, and those of degree $k+\dim M$ determine
certain differential $k$-forms on the space of connections $\mathcal{A}$ on $%
P$.

The equivariant characteristic forms provide canonical equivariant
extensions of these forms, and therefore canonical cohomology classes on $%
\mathcal{A}/\mathrm{Gau}^{0}P$. More generally, for any closed $\beta \in
\Omega ^{r}(M)$ and $f\in \mathcal{I}_{k}^{G}$, with $2k+r\geq \dim M$, a
cohomology class on $\mathcal{A}/\mathrm{Gau}^{0}P$ is obtained. These
classes are shown to coincide with some classes previously defined by Atiyah
and Singer.
\end{abstract}

\bigskip


\smallskip


\smallskip


\smallskip 

\noindent \emph{Acknowledgements.} Supported by the Ministerio of Ciencia y
Tecnolog\'{i}a of Spain, under grant \#BFM2002--00141.

\section{Introduction\label{intro}}

Let $\pi \colon P\rightarrow M$, be a principal $G$-bundle and let $p\colon
C(P)\rightarrow M$ be its bundle of connections. Let $\mathcal{I}_{k}^{G}$
be the space of Weil polynomials of degree $k$ for $G$. The principal $G$%
-bundle $C(P)\times _{M}P\rightarrow C(P)$ is endowed with a canonical
connection $\mathbb{A}$ (see below for the details), which can be used to
obtain, for every $f\in \mathcal{I}_{k}^{G}$, a characteristic $2k$-form on $%
C(P)$, denoted by $c_{f}(\mathbb{F})=f(\mathbb{F},\overset{(k}{\ldots },%
\mathbb{F})$ (\emph{e.g.}, see \cite{conn2}), where $\mathbb{F}$ is the
curvature of $\mathbb{A}$. Moreover, such a form is closed and $\mathrm{Aut}P
$-invariant. As $C(P)$ is an affine bundle, the map $p^{\ast }\colon H^{\ast
}(M)\rightarrow H^{\ast }(C(P))$ is an isomorphism. The cohomology class in $%
M$ corresponding to $c_{f}(\mathbb{F})$ under this isomorphism is the
characteristic class of $P$ associated to $f$. Hence, the characteristic
forms on $C(P)$ determine the characteristic classes on $M$, but the
characteristic forms contain more information than the characteristic
classes; for example, the characteristic classes of degree $2k>n$ vanish,
although the corresponding forms do not necessarily, as\emph{\ }$\dim
C(P)>\dim M$. Precisely, the principal aim of this paper is to provide a
geometric interpretation of such characteristic forms of higher degree.

This is based on the following construction. Let $E\rightarrow N$ be an
arbitrary bundle over a compact, oriented $n$-manifold without boundary. We
define a map $\digamma \colon \Omega ^{n+k}(J^{r}E)\rightarrow \Omega
^{k}(\Gamma (E))$ commuting with the exterior differential and with the
action of the group $\mathrm{Proj}^{+}(E)$ of projectable diffeomorphisms
which preserve the orientation on $M$. Hence, if $\alpha \in \Omega
^{n+k}(J^{r}E)$ is closed, exact, or invariant under a subgroup $\mathcal{G}%
\subset \mathrm{Proj}^{+}(E)$, then the form $\digamma \lbrack \alpha ]$
enjoys the same property.

Applying this construction to the bundle $C(P)\rightarrow M$, for any
characteristic form $c_{f}(\mathbb{F})$ with $2k>n$, we obtain a closed and $%
\mathrm{Gau}P$-invariant $(2k-n)$-form on the space $\mathcal{A}=\Gamma
(M,C(P))$ of connections on $P$. More generally, as proved in \cite{conn2},
the space of $\mathrm{Gau}P$-invariant forms on $C(P)$ is generated by forms
of type $c_{f}(\mathbb{F})\wedge p^{\ast }\beta $, with $\beta \in \Omega
^{\ast }(M)$. So, given $f\in \mathcal{I}_{k}^{G}$ and a closed $\beta \in
\Omega ^{r}(M)$, such that $2k+r\geq n$, we have a closed and $\mathrm{Gau}P$%
-invariant $(2k+r-n)$-form on $\mathcal{A}$ given by, 
\begin{equation}
C_{f,\beta }=\digamma \lbrack c_{f}(\mathbb{F})\wedge p^{\ast }\beta ]\in
\Omega ^{2k+r-n}(\mathcal{A}).  \label{form}
\end{equation}
As $\mathcal{A}$\ is an affine space, these forms are exact, and the
cohomology classes defined by them on $\mathcal{A}$, vanish; but in gauge
theories---because of gauge symmetry---it is more interesting to consider
the quotient space $\mathcal{A}/\mathrm{Gau}P$ instead of the space $%
\mathcal{A}$ itself. Although the forms (\ref{form}) are $\mathrm{Gau}P$%
-invariant, they are not projectable with respect to the natural quotient
map $\mathcal{A}\rightarrow \mathcal{A}/\mathrm{Gau}P$. Hence they do not
define directly cohomology classes on $\mathcal{A}/\mathrm{Gau}P$.
Consequently, we are led to consider another way in order to obtain
cohomology classes on the quotient from these forms. As is well known, the
cohomology of the quotient manifold by the action of a Lie group, is related
to the equivariant cohomology of the manifold; \emph{e.g.},\emph{\ }see \cite
{GS}. Below, we show that the usual construction of equivariant
characteristic classes (\emph{e.g.}, see \cite{BV1,BV2,BT}) when applied to
the canonical connection $\mathbb{A}$, provides canonical $\mathrm{Aut}P$%
-equivariant extensions of the characteristic forms. By extending the map $%
\digamma $ to equivariant differential forms in an obvious way, this result
allows us to obtain $\mathrm{Gau}P$-equivariant extensions of the forms (\ref
{form}); see Theorem \ref{equiext} below. These extensions determine
cohomology classes in the quotient space $\mathcal{A}/\mathrm{Gau}^{0}P$,
where $\mathrm{Gau}^{0}P\subset \mathrm{Gau}P$ is the subgroup of gauge
transformations preserving a fixed point $u_{0}\in P$. We also prove that
such classes coincide with those defined in \cite{AS}.

As is well known (\emph{e.g. }see \cite{AB2}), an equivariant extension of
an invariant symplectic two-form is equivalent to a moment map for it.
Hence, if the form (\ref{form}) is of degree two on $\mathcal{A}$, then the $%
\mathrm{Gau}P$-equivariant extension that we obtain, defines a canonical
moment map for the symplectic action of the gauge group on $\mathcal{A}$,
and\ we show that this symplectic forms and moment maps coincide with those
defined in \cite{AB1,Donaldson,Leung}.

Finally we show how our constructions lead to conservation laws for the
Chern-Simons terms considered in \cite{GMS}.

\section{The bundle of connections and the canonical connection\label{sec1}}

If $\pi \colon P\rightarrow M$ is a principal $G$-bundle, its bundle of
connections is an affine bundle $p\colon C(P)\rightarrow M$ modelled over
the vector bundle $T^{\ast }M\otimes \mathrm{ad}P$, such that there is a
bijection between connections on $P$ and the sections of $C(P)$ (\emph{e.g.}
see \cite{conn, gar,MS}). The natural projection $\bar{p}\colon \mathbb{P}%
=C(P)\times _{M}P\rightarrow C(P)$ onto the first factor induces a principal 
$G$-bundle structure over $C(P)$, and we have the commutative diagram 
\begin{equation*}
\begin{array}{ccc}
\mathbb{P} & \overset{\bar{p}}{\longrightarrow } & P \\ 
{\scriptstyle\bar{\pi}}\downarrow &  & \downarrow {\scriptstyle\pi } \\ 
C(P) & \overset{p}{\longrightarrow } & M
\end{array}
\end{equation*}
The bundle $\mathbb{P}$ has a canonical connection $\mathbb{A}\in \Omega
^{1}(\mathbb{P},\frak{g})$ characterized by, 
\begin{equation}
\mathbb{A}_{(\sigma _{A}(x),u)}(X)=A_{u}(\bar{p}_{\ast }X),
\label{defthetaA}
\end{equation}
for every connection $A$ on $P$, $x\in M$, $u\in \pi ^{-1}(x)$, $X\in
T_{(\sigma _{A}(x),u)}\mathbb{P}$, and where $\sigma _{A}\colon M\rightarrow
C(P)$ is the section corresponding to $A$.

\begin{remark}
\emph{It can be shown (see \cite{conn}) that the bundle }$\bar{p}\colon
P\rightarrow C(P)$\emph{\ is isomorphic to }$J^{1}P\rightarrow (J^{1}P)/G$%
\emph{\ and, under this identification, the canonical connection }$\mathbb{A}
$\emph{\ corresponds to the structure form of }$J^{1}P$\emph{.}
\end{remark}

The canonical connection enjoys the following properties (\emph{e.g.} see 
\cite{conn}):

\begin{enumerate}
\item[(1)]  $\mathbb{A}$ is invariant under the natural action of the group $%
\mathrm{Aut}P$ of automorphisms of $P$.

\item[(2)]  For every connection $A$ on $P$, we have $\bar{\sigma}_{A}^{\ast
}(\mathbb{A})=A$, where $\bar{\sigma}_{A}\colon P\rightarrow \mathbb{P}$ is
defined by $\bar{\sigma}_{A}(u)=(\sigma _{A}(x),u)$, with $x\in M$, $u\in
\pi ^{-1}(x)$.
\end{enumerate}

Let $\mathbb{F}$ be the curvature of $\mathbb{A}$. If $f\in \mathcal{I}%
_{k}^{G}$ is a Weil polynomial of degree $k$ for $G$, we define the
characteristic form associated to $f$ as the $2k$-form on $C(P)$ defined by $%
c_{f}(\mathbb{F})=f(\mathbb{F},\ldots ,\mathbb{F})$. This form has the
following properties:

\begin{enumerate}
\item[(3)]  $c_{f}(\mathbb{F})$ is closed.

\item[(4)]  $c_{f}(\mathbb{F})$ is invariant under the action of the group $%
\mathrm{Aut}P$ on $C(P)$.

\item[(5)]  For every connection $A$ on $P$ we have $\sigma _{A}^{\ast
}(c_{f}(\mathbb{F}))=f(F_{A},\ldots ,F_{A})$.
\end{enumerate}

As a consequence of (3) and (5) and the fact that the space of connections
is an affine space, we obtain the well-known result of Chern-Weil theory
that the cohomology class $[f(F_{A},\ldots ,F_{A})]\in H^{2k}(M)$ is
independent of the connection $A$, and is it called the characteristic class
associated to $f$. In other words, the map $\sigma _{A}^{\ast }$ is an
inverse of $p^{\ast }\colon H^{\bullet }(M)\rightarrow H^{\bullet }(C(P))$,
and under this isomorphism the cohomology class of $c_{f}(\mathbb{F})$
corresponds to the characteristic class of$\ P$ associated to $f$ (\emph{e.g.%
} see \cite{conn2,MS}).

The space of connections $\mathcal{A}$ is an affine space modelled over $%
\Omega ^{1}(M,\mathrm{ad}P)$. Hence, we have the identification $T_{A}%
\mathcal{A}\simeq \Omega ^{1}(M,\mathrm{ad}P)$ for every $A\in \mathcal{A}$.
Also, $C(P)$ is an affine bundle modelled over the vector bundle $T^{\ast
}M\otimes \mathrm{ad}P$. So, for every $a\in \Omega ^{1}(M,\mathrm{ad}%
P)\subset \Gamma (C(P),T^{\ast }M\otimes \mathrm{ad}P)$ we have a vertical
vector field $X_{a}\in \frak{X}^{v}(C(P))$.

\begin{lemma}
\label{Xa}For every $a,b\in \Omega ^{1}(M,\mathrm{ad}P)$, we have 
\begin{eqnarray*}
i_{X_{a}}\mathbb{F} &=&p^{\ast }a, \\
i_{X_{b}}i_{X_{a}}\mathbb{F} &=&0.
\end{eqnarray*}
\end{lemma}

\begin{proof}
It follows from the formula (5.8) in \cite{conn}.
\end{proof}

If $A_{0}$, $A_{1}\in \mathcal{A}$, define $A_{t}=(1-t)A_{0}+tA_{1}$ and $%
a=A_{1}-A_{0}\in \Omega ^{1}(M,\mathrm{ad}P)$. The tangent vector to the
curve $\sigma _{A_{t}}(x)$ in $C(P)$ is $X_{a}(\sigma _{A_{t}}(x))$ for any $%
x\in M$, and hence we recover the usual transgression formula 
\begin{equation*}
c_{f}(F_{A_{1}})\!-\!c_{f}(F_{A_{0}})\!=\!d\left( \!\int_{0}^{1}\sigma
_{A_{t}}^{\ast }\left( i_{X_{a}}c_{f}(\mathbb{F})\right) dt\!\right)
\!=\!d\left( \!k\int_{0}^{1}f(a,F_{A_{t}},\ldots ,F_{A_{t}})dt\!\right) .
\end{equation*}

Given a connection $A_{0}$ on $P$, $\bar{p}^{\ast }A_{0}$ is a connection on 
$\mathbb{P}$. As $\bar{p}^{\ast }A_{0}$ and $\mathbb{A}$ are connections on
the same bundle, defining $a_{0}=\mathbb{A}-\bar{p}^{\ast }A_{0}\in \Omega
^{1}(C(P),\frak{g})$, $A_{t}=(1-t)\bar{p}^{\ast }A_{0}+t\mathbb{A}$ and 
\begin{equation*}
\eta _{f}(A_{0})=k\int_{0}^{1}f(a_{0},F_{A_{t}},\ldots ,F_{A_{t}})dt,
\end{equation*}
we have $c_{f}(\mathbb{F})-c_{f}(F_{\bar{p}^{\ast }A_{0}})=d\eta _{f}(A_{0})$%
. If $2k>n$, then $c_{f}(F_{\bar{p}^{\ast }A_{0}})=p^{\ast
}c_{f}(F_{A_{0}})=0$, and hence $c_{f}(\mathbb{F})=d\eta _{f}(A_{0})$.

\section{Equivariant Characteristic forms\label{sec1.2}}

First, we recall the definition of equivariant cohomology in the Cartan
model (\emph{e.g. }see \cite{BGV,GS}). Suppose that we have a left action of
a connected Lie group $\mathcal{G}$ on a manifold $N$, \emph{i.e.} a
homomorphism $\rho \colon \mathcal{G}\rightarrow \mathrm{Diff}(N)$. We have
an induced Lie algebra homomorphism 
\begin{eqnarray*}
\mathrm{Lie\,}\mathcal{G} &\rightarrow &\frak{X}(N) \\
X &\mapsto &X_{N}=\left. \frac{d}{dt}\right| _{t=0}\rho (\exp (-tX))
\end{eqnarray*}
Let $\Omega _{\mathcal{G}}(N)=\left( \mathbf{S}^{\bullet }(\mathrm{Lie\,}%
\mathcal{G}^{\ast })\otimes \Omega ^{\bullet }(N)\right) ^{\mathcal{G}}=%
\mathcal{P}^{\bullet }(\mathrm{Lie\,}\mathcal{G},\Omega ^{\bullet }(N))^{%
\mathcal{G}}$ be the space of $\mathcal{G}$-invariant polynomials on $%
\mathrm{Lie\,}\mathcal{G}$ with values in $\Omega ^{\bullet }(N)$. We define
the following graduation: $\deg (\alpha )=2k+r$ if $\alpha \in \mathcal{P}%
^{k}(\mathrm{Lie\,}\mathcal{G},\Omega ^{r}(N))$. Hence the space of $%
\mathcal{G}$-equivariant differential $q$-forms is 
\begin{equation*}
\Omega _{\mathcal{G}}^{q}(N)=\bigoplus_{2k+r=q}(\mathcal{P}^{k}(\mathrm{Lie\,%
}\mathcal{G},\Omega ^{r}(N)))^{\mathcal{G}}.
\end{equation*}
Let $d_{c}\colon \Omega _{\mathcal{G}}^{q}(N)\rightarrow \Omega _{\mathcal{G}%
}^{q+1}(N)$ be the Cartan differential, 
\begin{equation*}
(d_{c}\alpha )(X)=d(\alpha (X))-i_{X_{N}}\alpha (X),\quad \forall X\in 
\mathrm{Lie\,}\mathcal{G}.
\end{equation*}
As is well known, on $\Omega _{\mathcal{G}}^{\bullet }(N)$ we have $\left(
d_{c}\right) ^{2}=0$. Moreover, the equivariant cohomology (in the Cartan
model) of $N$ with respect of the action of $\mathcal{G}$ is defined as the
cohomology of this complex, \emph{i.e.}, 
\begin{equation*}
H_{\mathcal{G}}^{q}(N)=\frac{\mathrm{\ker }\left( d_{c}\colon \Omega _{%
\mathcal{G}}^{q}(N)\rightarrow \Omega _{\mathcal{G}}^{q+1}(N)\right) }{%
\mathrm{Im}\left( d_{c}\colon \Omega _{\mathcal{G}}^{q-1}(N)\rightarrow
\Omega _{\mathcal{G}}^{q}(N)\right) }.
\end{equation*}

\begin{definition}
\emph{Given a closed and }$\mathcal{G}$\emph{-invariant form }$\omega \in
\Omega ^{q}(M)$\emph{, an equivariant differential form }$\omega ^{\#}\in
\Omega _{\mathcal{G}}^{q}(M)$\emph{\ is said to be a }$\mathcal{G}$\emph{%
-equivariant extension of }$\omega $\emph{\ if }$d_{c}\omega ^{\#}=0$\emph{\
and }$\omega ^{\#}(0)=\omega $\emph{.}
\end{definition}

In general, there could be obstructions to the existence of equivariant
extensions (\emph{e.g.},\emph{\ }see \cite{Wu}) but, as we will see, the
classical construction of equivariant characteristic classes really provides
canonical equivariant extensions for the forms we are dealing with.

Next, let us recall the relationship between equivariant cohomology and the
cohomology of the quotient space.\ If the action of $\mathcal{G}$ on $N$ is
free and $N/\mathcal{G}$ is a manifold, then $N\rightarrow N/\mathcal{G}$ is
a principal $\mathcal{G}$-bundle. Let $A$ be a connection on this bundle.
The following map is a generalization of the Chern-Weil homomorphism: 
\begin{eqnarray*}
\mathrm{ChW}_{A}\colon \Omega _{\mathcal{G}}^{\bullet }(N) &\rightarrow
&\left( \Omega ^{\bullet }(N)\right) _{\mathrm{basic}}\simeq \Omega
^{\bullet }(N/\mathcal{G}) \\
\alpha &\mapsto &\left( \alpha (F_{A})\right) _{\mathrm{hor}}
\end{eqnarray*}
where $\beta _{\mathrm{hor}}$ denotes the horizontal component of $\beta \in
\Omega ^{\bullet }(N)$ with respect to the connection $A$. We have

\begin{proposition}
\label{ddc}If $\alpha \in \Omega _{\mathcal{G}}^{\bullet }(N)$, then $%
\mathrm{ChW}_{A}(d_{c}\alpha )=d(\mathrm{ChW}_{A}(\alpha ))$.
\end{proposition}

\begin{proof}
We refer the reader to \cite[Theorem 7.34]{BGV}.
\end{proof}

\begin{theorem}
\label{equcoc}The induced map in cohomology $\mathrm{ChW}_{A}\colon H_{%
\mathcal{G}}^{\bullet }(N)\rightarrow H^{\bullet }(N/\mathcal{G})$ is
independent of the connection $A$ chosen, and is denoted by 
\begin{equation*}
\mathrm{ChW}_{N}\colon H_{\mathcal{G}}^{\bullet }(N)\rightarrow H^{\bullet
}(N/\mathcal{G}).
\end{equation*}
\end{theorem}

\begin{proof}
The result quickly follows by working on the bundle of connections. We use
the notations introduced in Section \ref{sec1} by setting $P=N$, $M=N/%
\mathcal{G}$ and denoting by $p\colon P\rightarrow M$ the quotient map. Let $%
\alpha \in \Omega _{\mathcal{G}}^{q}(P)$ be an equivariant $q$-form such
that $d_{c}\alpha =0$. Recall that $\bar{p}^{\ast }(\alpha )$ belongs to $%
\Omega _{\mathcal{G}}^{q}(\mathbb{P})$ as $\bar{p}$ is a $\mathcal{G}$%
-equivariant map. By Proposition \ref{ddc}, the form $\mathrm{ChW}_{\mathbb{A%
}}(\bar{p}^{\ast }\alpha )\in \Omega ^{\bullet }(C(P))$ is closed, and from
the formula (\ref{defthetaA}) we obtain 
\begin{equation*}
\sigma _{A}^{\ast }(\mathrm{ChW}_{\mathbb{A}}(\bar{p}^{\ast }\alpha ))=%
\mathrm{ChW}_{A}(\alpha ).
\end{equation*}
Again the result follows as the space of connections is contractible.
\end{proof}

\begin{remark}
\emph{If }$\mathcal{G}$ \emph{is compact and connected} $\mathrm{ChW}_{N}$ 
\emph{is an isomorphism (}e.g.\emph{\ see \cite{GS}).}
\end{remark}

The definition of equivariant characteristic classes of Berline and Vergne
(see \cite{BV1,BV2,BT}) can be introduced as follows. Let $\pi \colon
P\rightarrow M$ a principal $G$-bundle and let us further assume that a Lie
group $\mathcal{G}$ acts (on the left) on $P$ by automorphisms of this
bundle. Let $A$ be a connection on $P$, which is \emph{invariant under the
action of }$\mathcal{G}$.

For every $f\in \mathcal{I}_{k}^{G}$ the $\mathcal{G}$-equivariant
characteristic form associated to $f$ and $A$, $c_{f}(F_{A}^{\mathcal{G}%
})\in \Omega _{\mathcal{G}}^{2k}(M)$, is defined by 
\begin{eqnarray*}
c_{f}(F_{A}^{\mathcal{G}})(X) &=&f\left( F_{A}-A(X_{P}),\overset{(k}{\ldots }%
,F_{A}-A(X_{P})\right) \\
&=&\sum_{i=1}^{k}(-1)^{k-i}\tbinom{k}{i}f(F_{A},\overset{(i}{\ldots }%
,F_{A},A(X_{P}),\overset{(k-i}{\ldots },A(X_{P}))
\end{eqnarray*}
for every $X\in \mathrm{Lie}\mathcal{G}$.

\begin{proposition}
\label{equiex}We have

\begin{enumerate}
\item[$(1)$]  $c_{f}(F_{A}^{\mathcal{G}})$ is a $\mathcal{G}$-equivariant
extension of $c_{f}(F_{A})$.

\item[$(2)$]  The equivariant cohomology class $c_{f}^{\mathcal{G}%
}(P)=[c_{f}(F_{A}^{\mathcal{G}})]\in H_{\mathcal{G}}^{2k}(M)$ is independent
of the $\mathcal{G}$-invariant connection $A$, and is called the $\mathcal{G}
$-equivariant cohomology class of $P\,$associated to $f$.
\end{enumerate}
\end{proposition}

\begin{proof}
See \cite{BT,BV2}).
\end{proof}

Applying the construction of equivariant characteristic forms to the bundle $%
\mathbb{P}\rightarrow C(P)$ with the $\mathrm{Aut}P$-invariant connection $%
\mathbb{A}$, we obtain the $\mathrm{Aut}P$-equivariant characteristic form $%
c_{f}(\mathbb{F}^{\mathrm{Aut}P})\in \Omega _{\mathrm{Aut}P}^{2k}(C(P))$,
with is an $\mathrm{Aut}P$-equivariant extension of $c_{f}(\mathbb{F})$. If $%
\mathcal{G}\subset \mathrm{Aut}P$ is any subgroup of the automorphism group,
we have the corresponding $\mathcal{G}$-equivariant characteristic form 
\begin{equation*}
c_{f}(\mathbb{F}^{\mathcal{G}})=c_{f}(\mathbb{F}^{\mathrm{Aut}P})|_{\mathrm{%
Lie}\mathcal{G}}.
\end{equation*}
The following Proposition easily follows from the formula (\ref{defthetaA}):

\begin{proposition}
\label{pbequi}If $\mathcal{G}$ acts on $\pi \colon P\rightarrow M$ by
automorphisms of $\pi $, and $A$ is a $\mathcal{G}$-invariant connection on $%
P$, then we have $\sigma _{A}^{\ast }(c_{f}(\mathbb{F}^{\mathcal{G}%
}))=c_{f}(F_{A}^{\mathcal{G}}).$
\end{proposition}

\begin{remark}
\emph{In this way, we obtain the analogous situation to that of the ordinary
characteristic classes; see the last paragraph in Section \ref{sec1}.
Moreover, Proposition \ref{pbequi} provides a very simple proof of
Proposition \ref{equiex}--}$(2)$\emph{, as the space of }$\mathcal{G}$\emph{%
-invariant connections is an affine subspace; more precisely, if }$A$\emph{\
is a }$\mathcal{G}$\emph{-invariant connection, }$\sigma _{A}^{\ast }$\emph{%
\ is the inverse of }$p^{\ast }\colon H_{\mathcal{G}}^{\bullet
}(M)\rightarrow H_{\mathcal{G}}^{\bullet }(C(P))$\emph{\ (hence }$p^{\ast }$%
\emph{\ is an isomorphism). Under this isomorphism the }$G$\emph{%
-equivariant cohomology class of }$c_{f}(\mathbb{F}^{\mathcal{G}})$ \emph{%
corresponds to the} $\mathcal{G}$\emph{-equivariant characteristic class
associated to} $f$\emph{. Moreover, as in the case of ordinary
characteristic classes, the equivariant characteristic forms contain more
information than their corresponding characteristic classes. For example, in
Section \ref{sec3} we will use this forms in the case} $\mathcal{G}=\mathrm{%
Gau}P$ \emph{to find equivariant extensions of the forms (\ref{form}).}
\end{remark}

The analog of Proposition \ref{equcoc} for the equivariant characteristic
classes, is

\begin{proposition}
\label{quotcha}Assume that $\mathcal{G}$ acts freely on $P$ and $M$, and
that the quotient bundle $\pi _{\mathcal{G}}\colon P/\mathcal{G}\rightarrow
M/\mathcal{G}$ exists, then 
\begin{equation*}
\mathrm{ChW}_{M}(c_{f}^{\mathcal{G}}(P))=c_{f}(P/\mathcal{G}).
\end{equation*}
\end{proposition}

\begin{proof}
We denote by $q_{P}\colon P\rightarrow P/\mathcal{G}$, $q_{M}\colon
M\rightarrow M/\mathcal{G}$ the projections. Let $A_{1}$ a connection on the
principal $G$-bundle $\pi _{\mathcal{G}}\colon P/\mathcal{G}\rightarrow M/%
\mathcal{G}$, and $A_{2}$ a connection in the principal $\mathcal{G}$-bundle 
$M\rightarrow M/\mathcal{G}$. Clearly $A_{1}^{\prime }=q_{P}^{\ast }A_{1}$
is a $\mathcal{G}$-invariant connection on the principal $G$-bundle $%
P\rightarrow M$, and for every $X\in \mathrm{Lie}\mathcal{G}$ we have $%
A_{1}^{\prime }(X_{P})=0$. So, the equivariant characteristic class
associated to $A_{1}^{\prime }$ and $f$ is the basic form $%
c_{f}(F_{A_{1}^{\prime }}^{\mathcal{G}})=c_{f}(F_{A_{1}^{\prime }})$.

From the very definition of $\mathrm{ChW}_{A_{2}}$, it is clear that 
\begin{equation*}
\mathrm{ChW}_{A_{2}}(c_{f}(F_{A_{1}^{\prime }}))=c_{f}(F_{A_{1}}),
\end{equation*}
and hence the result follows.
\end{proof}

\section{Forms in $\Gamma (E)$ induced by forms in $J^{r}E$\label{sec2}}

Let $q\colon E\rightarrow M$ be a locally trivial bundle over an oriented,
connected, and compact $n$-manifold without boundary $M$. We denote by $%
\Gamma (E)$ the space of global sections of $E$, and we assume that it is
not empty. We consider $\Gamma (E)$ as a differential manifold; for the
details of its infinite-dimensional structure see \cite{KM}. For any $s\in
\Gamma (E)$ there is an identification $T_{s}\Gamma (E)\simeq \Gamma
(M,s^{\ast }V(E))$. We denote by $J^{r}E$ the $r$-jet bundle of $E$, and by $%
\mathrm{Proj}(E)$ the group of projectable diffeomorphisms of $E$, \emph{i.e.%
} $\phi \in \mathrm{Diff}(E)$ such that there exist $\underline{\phi }\in 
\mathrm{Diff}(M)$ with $q\circ \underline{\phi }=\phi \circ q$. We denote by 
$\mathrm{Proj}^{+}(E)$ the subgroup of elements $\phi \in \mathrm{Proj}(E)$
such that $\underline{\phi }\in \mathrm{Diff}^{+}(M)$, the group of
orientation preserving diffeomorphisms.

We denote by $\mathrm{proj}(E)\subset \frak{X}(E)$ the Lie algebra of
projectable vector fields, which can be considered as the Lie algebra of $%
\mathrm{Proj}(E)$. The group $\mathrm{Proj}(E)$ acts on $\Gamma (E)$ by $%
(\phi ,s)\mapsto \phi _{\Gamma (E)}(s)=\phi \circ s\circ \underline{\phi }%
^{-1}$. At the Lie-algebra level, every projectable vector field $X\in 
\mathrm{proj}(E)$ determines a vector field $X_{\Gamma (E)}\in \frak{X}%
(\Gamma (E))$ on $\Gamma (E)$. Given $\phi \in $ $\mathrm{Proj}(E)$ (resp. $%
X\in \mathrm{proj}(E)$) we denote by $\phi ^{(r)}$ (resp. $X^{(r)}$) its
prolongation to $J^{r}(E).$ We recall that $\phi ^{(r)}(j_{x}^{r}s)=j_{%
\underline{\phi }(x)}^{r}\left( \phi _{\Gamma (E)}(s)\right) $.

The evaluation map 
\begin{eqnarray*}
\mathrm{ev}_{r}\colon M\times \Gamma (E) &\rightarrow &J^{r}E \\
(x,s) &\mapsto &j_{x}^{r}s
\end{eqnarray*}
is equivariant with respect of the action of $\mathrm{Proj}(E)$ on $M\times
\Gamma (E)$ and $J^{r}E$. So, for any $X\in \mathrm{proj}(E)$, denoting by $%
\underline{X}\in \frak{X}(E)$ its projection to $M$, we have 
\begin{equation}
\mathrm{ev}_{r\ast }(\underline{X},X_{\Gamma (E)})=X^{(r)}.  \label{evX}
\end{equation}

We define a map 
\begin{equation*}
\digamma \colon \Omega ^{n+k}(J^{r}E)\rightarrow \Omega ^{k}(\Gamma (E))
\end{equation*}
by the formula 
\begin{equation}
\digamma \lbrack \alpha ]=\int_{M}\mathrm{ev}_{r}^{\ast }\alpha \in \Omega
^{k}(\Gamma (E)),  \label{ev}
\end{equation}
where $\int_{M}$ denotes the integration over the fiber of $M\times \Gamma
(E)\rightarrow \Gamma (E)$. If $\alpha \in \Omega ^{k}(J^{r}E)$ with $k<n$%
,we set $\digamma \lbrack \alpha ]=0$.

\begin{proposition}
For any $\alpha \in \Omega ^{n+k}(J^{r}E)$, we have 
\begin{equation}
(\digamma \lbrack \alpha ])_{s}(X_{1},\ldots ,X_{k})=\int_{M}\left(
j^{r}s\right) ^{\ast }(i_{X_{k}^{(r)}}\ldots i_{X_{1}^{(r)}}\alpha ),
\label{defF}
\end{equation}
for every $s\in \Gamma (E)$, $X_{1},\ldots ,X_{k}\in T_{s}\Gamma E\simeq
\Gamma (M,s^{\ast }V(E)).$
\end{proposition}

\begin{proof}
The result follows from the definition of $\digamma \lbrack \alpha ]$ and
the formula (\ref{evX}) applied to vertical vector fields.
\end{proof}

The following proposition follows from the definition of $\digamma $ and the
properties of the integration over the fiber

\begin{proposition}
\label{propF}For every $\alpha \in \Omega ^{n+k}(J^{r}E)$, $\phi \in \mathrm{%
Proj}^{+}(E)$, and $X\in \mathrm{proj}(E)$ we have

\begin{enumerate}
\item[\emph{a)}]  $\digamma \lbrack d\alpha ]=\mathrm{d}\digamma \lbrack
\alpha ],$

\item[\emph{b)}]  $\digamma \lbrack \left( \phi ^{(r)}\right) ^{\ast }\alpha
]=\phi _{\Gamma (E)}^{\ast }\digamma \lbrack \alpha ],$

\item[\emph{c)}]  $\digamma \lbrack i_{X^{(r)}}\alpha ]=\mathrm{i}%
_{X_{\Gamma (E)}}\digamma \lbrack \alpha ],$

\item[\emph{d)}]  $\digamma \lbrack L_{X^{(r)}}\alpha ]=\mathrm{L}%
_{X_{\Gamma (E)}}\digamma \lbrack \alpha ].$
\end{enumerate}
\end{proposition}

\begin{remark}
\emph{If }$\alpha \in \Omega ^{n-1}(J^{r}E)$\emph{, the condition \ref{propF}%
--a) means} $\digamma \lbrack d\alpha ]=0$\emph{.}
\end{remark}

Now, assume that $\mathcal{G}$ is a subgroup of $\mathrm{Proj}^{+}(E)$. If $%
\alpha \in \mathcal{P}^{q}(\mathrm{Lie\,}\mathcal{G},\Omega ^{n+k}(E))$, the
composition 
\begin{equation*}
\mathrm{Lie\,}\mathcal{G}\overset{\alpha }{\longrightarrow }\Omega
^{n+k}(J^{r}E)\overset{\digamma }{\longrightarrow }\Omega ^{k}(\Gamma (E))
\end{equation*}
defines an element $\digamma \lbrack \alpha ]$ of $\mathcal{P}^{q}(\mathrm{%
Lie\,}\mathcal{G},\Omega ^{k}(\Gamma (E)))$; that is, for $X\in \mathrm{Lie\,%
}\mathcal{G}$ we have, 
\begin{equation*}
\left( \digamma \lbrack \alpha ]\right) (X)=\digamma \lbrack \alpha (X)].
\end{equation*}
By Proposition \ref{propF}-b) if $\alpha $ is $\mathcal{G}$-invariant, $%
\digamma \lbrack \alpha ]$ is also $\mathcal{G}$-invariant, and so the map $%
\digamma $\ extend to a map between $\mathcal{G}$-equivariant differential
forms, 
\begin{equation*}
\digamma \colon \Omega _{\mathcal{G}}^{n+k}(J^{r}E)\rightarrow \Omega _{%
\mathcal{G}}^{k}(\Gamma (E)).
\end{equation*}

\begin{proposition}
For every $\alpha \in \Omega _{\mathcal{G}}^{n+k}(J^{r}E)$ we have $\digamma
\lbrack d_{c}\alpha ]=\mathrm{d}_{c}\digamma \lbrack \alpha ]$. Hence, we
have an induced map in equivariant cohomology $\digamma \colon H_{\mathcal{G}%
}^{n+k}(J^{r}E)\rightarrow H_{\mathcal{G}}^{k}(\Gamma (E)).$
\end{proposition}

\begin{proof}
If $\alpha \in \Omega _{\mathcal{G}}^{n+k}(J^{r}E)$ and $X\in \mathrm{Lie\,}%
\mathcal{G}$, then from Proposition \ref{propF} we have 
\begin{eqnarray*}
\left( \digamma \lbrack d_{c}\alpha ]\right) (X) &=&\digamma \lbrack
d_{c}\alpha (X)]=\digamma \lbrack d(\alpha (X))]-\digamma \lbrack
i_{X^{(r)}}\alpha (X)] \\
&=&\mathrm{d}\digamma \lbrack \alpha (X)]-\mathrm{i}_{X_{\Gamma
(E)}}\digamma \lbrack \alpha (X)]=\left( \mathrm{d}_{c}\digamma \lbrack
\alpha ]\right) (X).
\end{eqnarray*}
\end{proof}

\section{Applications\label{sec3}}

In this section we combine the results of Sections \ref{sec1.2} and \ref
{sec2}. As remarked in the Introduction, in Gauge theories $\mathrm{Gau}P$%
-invariant forms are specially interesting, so we focus on these forms. In 
\cite{conn2} it is proved that the space of $\mathrm{Gau}P$-invariant forms
is generated by the forms of type $c_{f}(\mathbb{F})\wedge p^{\ast }\beta $,
with $f\in \mathcal{I}_{k}^{G}$ and $\beta \in \Omega ^{r}(M)$. We assume
that $\beta $ is closed and $2k+r\geq n$.

By applying the map $\digamma $ to $c_{f}(\mathbb{F})\wedge p^{\ast }\beta $
we obtain 
\begin{equation*}
C_{f,\beta }=\digamma \lbrack c_{f}(\mathbb{F})\wedge p^{\ast }\beta ]\in
\Omega ^{2k+r-n}(\mathcal{A}).
\end{equation*}
By Proposition \ref{propF} this form is closed and $\mathrm{Gau}P$-invariant.

Taking Lemma \ref{Xa}\ into account, it si easy to obtain the expression of $%
C_{f,\beta }$. We have

\begin{proposition}
\label{expromega}Let $q=2k+r-n$. For $a_{1},\ldots ,a_{q}\in \Omega ^{1}(M,%
\mathrm{ad}P)$ we have: 
\begin{equation*}
(C_{f,\beta })_{A}(a_{1},\ldots ,a_{q})=\tbinom{k}{q}\int_{M}f(a_{1},\ldots
,a_{q},F_{A},\overset{(n-k-r}{\ldots \ldots },F_{A})\wedge \beta
\end{equation*}
\end{proposition}

\noindent%
%
By virtue of Proposition \ref{equiex}, the $\mathrm{Gau}P$-equivariant
characteristic form $c_{f}(\mathbb{F}^{\mathrm{Gau}P})$ is an equivariant
extension of $c_{f}(\mathbb{F})$. Also $p^{\ast }\beta $ is a closed $%
\mathrm{Gau}P$-equivariant differential form, because it is closed and
basic. So $c_{f}(\mathbb{F}^{\mathrm{Gau}P})\wedge p^{\ast }\beta $ is a $%
\mathrm{Gau}P$-equivariant extension of $c_{f}(\mathbb{F})\wedge p^{\ast
}\beta $. We thus obtain the following

\begin{theorem}
\label{equiext}The $\mathrm{Gau}P$-equivariant form 
\begin{equation}
C_{f,\beta }^{\#}=\digamma \lbrack c_{f}(\mathbb{F}^{\mathrm{Gau}P})\wedge
p^{\ast }\beta ]\in \Omega _{\mathrm{Gau}P}^{2k+r-n}(\mathcal{A})
\label{equiextomega}
\end{equation}
is a $\mathrm{Gau}P$-equivariant extension of $C_{f,\beta }$.
\end{theorem}

We have thus found a canonical $\mathrm{Gau}P$-equivariant extension of $%
C_{f,\beta }$, as said in the Introduction.

Let $X\in \mathrm{gau}P=\Omega ^{0}(M,\mathrm{ad}P)\simeq \Omega _{\mathrm{Ad%
}}^{0}(P,\frak{g})$, and $X_{\mathbb{P}}\in \frak{X(}\mathbb{P)}$ the vector
field corresponding to the action of $\mathrm{Gau}P$ in $\mathbb{P}$. We
have 
\begin{equation*}
\mathbb{A}_{(\sigma _{A}(x),u)}(X_{\mathbb{P}})=A_{u}(\bar{p}_{\ast }X_{%
\mathbb{P}})=A_{u}(X_{P})=X(u).
\end{equation*}
So, $\mathbb{A}(X_{\mathbb{P}})=p^{\ast }X$, and hence 
\begin{eqnarray}
c_{f}(\mathbb{F}^{\mathrm{Gau}P})(X) &=&f\left( \mathbb{F}-p^{\ast }X,%
\overset{(k}{\ldots },\mathbb{F}-p^{\ast }X\right)  \notag \\
&=&\sum_{i=0}^{k}c_{f}^{i}(\mathbb{F},X),  \label{expequi}
\end{eqnarray}
where $c_{f}^{i}(\mathbb{F},X)=(-1)^{i}\tbinom{k}{i}f(\mathbb{F},\overset{%
(k-i}{\ldots },\mathbb{F},p^{\ast }X,\overset{(i}{\ldots },p^{\ast }X)$.

The condition $d_{c}c_{f}(\mathbb{F}^{\mathrm{Gau}P})=0$ is equivalent to 
\begin{eqnarray}
dc_{f}(\mathbb{F}) &=&0,  \notag \\
dc_{f}^{i}(\mathbb{F},X) &=&i_{X_{C(P)}}c_{f}^{i-1}(\mathbb{F},X),\quad
i=1,\ldots ,k.  \label{descenso}
\end{eqnarray}

Using (\ref{expequi}) and Proposition \ref{expromega} it is easy to obtain
the expression for $C_{f,\beta }^{\#}$. In the Example \ref{simpectica} of
subsection \ref{subsec3.2}, we detail such expression in a simple case.
Also, for $f(X)=\mathrm{Tr}(\mathrm{\exp }X)$ we obtain the equivariant
forms defined in \cite[sec. 6]{Leung} as a particular case.

\subsection{Forms in $\mathcal{A}/\mathrm{Gau}^{0}P$\label{subsec3.1}}

Let $\mathrm{Gau}^{0}P\subset \mathrm{Gau}P$ be the subgroup of gauge
transformations acting as the identity on the fiber over a fixed point $%
x_{0}\in M$. Then $\mathrm{Gau}^{0}P$ acts freely on $\mathcal{A}$ and the
quotient $\mathcal{A}/\mathrm{Gau}^{0}P$ is well defined (\emph{e.g. }see 
\cite{MV,CRR}).

By virtue of Theorem \ref{equcoc}, the $\mathrm{Gau}^{0}P$-equivariant
differential form $C_{f,\beta }^{\#}$ determines a cohomology class 
\begin{equation*}
\mathrm{ChW}_{\mathcal{A}}(C_{f,\beta }^{\#})\in H^{2k+r-n}(\mathcal{A}/%
\mathrm{Gau}^{0}P).
\end{equation*}
To obtain a representative of this class it is necessary a connection on the
bundle $\mathcal{A}\rightarrow \mathcal{A}/\mathrm{Gau}^{0}P$. The
construction of this connection is a standard fact in gauge theories (\emph{%
e.g.} see \cite{CRR}): Given a Riemannian metric $g$ in $M$ there is a
connection in $\mathcal{A}\rightarrow \mathcal{A}/\mathrm{Gau}^{0}P$ given
by the decomposition 
\begin{equation*}
T_{A}\mathcal{A}\simeq \Omega ^{1}(M,\mathrm{ad}P)\simeq \mathrm{Im}%
(d_{A})\oplus \ker (d_{A}^{\ast }).
\end{equation*}
The space $\mathrm{Im}(d_{A})$ is the tangent space to the orbits, and $\ker
(d_{A}^{\ast })$ is the horizontal complement. The expression of its
corresponding connection form is 
\begin{equation}
\frak{A}=G_{A}d_{A}^{\ast },  \label{conA}
\end{equation}
where $G_{A}=(d_{A}^{\ast }\circ d_{A})^{-1}$ is the Green function of the
Laplacian, 
\begin{equation*}
\Delta _{A}^{0}=d_{A}^{\ast }\circ d_{A}\colon \Omega ^{0}(M,\mathrm{ad}%
P)\rightarrow \Omega ^{0}(M,\mathrm{ad}P).
\end{equation*}
We denote by $\frak{F}$ the curvature of $\frak{A}$.

Next, we relate our classes $\mathrm{ChW}_{\mathcal{A}}(C_{f,\beta }^{\#})$
to the constructions in \cite{AS}. Consider the principal $G$-bundle $%
P\times \mathcal{A}\rightarrow M\times \mathcal{A}$.\ The group $\mathrm{Gau}%
^{0}P$ acts on $P$ and on $\mathcal{A}$. Taking the quotient, we obtain a
principal $G$-bundle 
\begin{equation*}
\mathcal{Q}=\left( P\times \mathcal{A}\right) /\mathrm{Gau}^{0}P\rightarrow
M\times (\mathcal{A}/\mathrm{Gau}^{0}P).
\end{equation*}
If $f\in \mathcal{I}_{k}^{G}$ and $[\beta ]\in H^{r}(M)$, we have a class 
\begin{equation*}
c_{f}(\mathcal{Q})\wedge \lbrack \beta ]\in H^{2k+r}(M\times (\mathcal{A}/%
\mathrm{Gau}^{0}P)).
\end{equation*}
Integrating over $M$ we obtain the class 
\begin{equation*}
\mu _{f}([\beta ])=\int_{M}c_{f}(\mathcal{Q})\wedge \lbrack \beta ]\in
H^{2k+r-n}(\mathcal{A}/\mathrm{Gau}^{0}P).
\end{equation*}

\begin{theorem}
For every $f\in \mathcal{I}_{k}^{G}$ and every closed $\beta \in \Omega
^{r}(M)$ with $2k+r\geq n$, we have 
\begin{equation*}
\mathrm{ChW}_{\mathcal{A}}\left( C_{f,\beta }^{\#}\right) =\mu _{f}([\beta
]).
\end{equation*}
\end{theorem}

\begin{proof}
The evaluation map extends to a morphism of principal $G$-bundles 
\begin{equation*}
\begin{array}{ccc}
P\times \mathcal{A} & \overset{\overline{\mathrm{ev}}}{\longrightarrow } & 
\mathbb{P} \\ 
\downarrow &  & \downarrow {\scriptstyle\bar{\pi}} \\ 
M\times \mathcal{A} & \overset{\mathrm{ev}}{\longrightarrow } & C(P)
\end{array}
\end{equation*}
where $\overline{\mathrm{ev}}(u,A)=(\sigma _{A}(x),u)$ for $u\in \pi
^{-1}(x) $. Hence, $\overline{\mathrm{ev}}^{\ast }(\mathbb{A})$ is a
connection on $P\times \mathcal{A}\rightarrow M\times \mathcal{A}$, $\mathrm{%
ev}^{\ast }(c_{f}(\mathbb{F}))$ gives its characteristic classes, and $%
\mathrm{ev}^{\ast }(c_{f}(\mathbb{F}^{\mathrm{Gau}^{0}P}))$, $f\in \mathcal{I%
}_{k}^{G}$, are its $\mathrm{Gau}^{0}P$-equivariant characteristic forms. By
Proposition \ref{quotcha} these equivariant characteristic forms determine
the characteristic classes of the quotient bundle $\mathcal{Q}\rightarrow
M\times (\mathcal{A}/\mathrm{Gau}^{0}P)$, \emph{i.e.}, we have 
\begin{equation*}
\mathrm{ChW}_{M\times \mathcal{A}}(\mathrm{ev}^{\ast }(c_{f}(\mathbb{F}^{%
\mathrm{Gau}^{0}P})))=c_{f}(\mathcal{Q}).
\end{equation*}
As the connection on $M\times \mathcal{A}\rightarrow M\times (\mathcal{A}/%
\mathrm{Gau}^{0}P)$ is given by the connection (\ref{conA}), the cohomology
class $\mu _{f}([\beta ])$, is represented by the form 
\begin{equation}
\left( \int_{M}c_{f}(\mathrm{ev}^{\ast }(\mathbb{F})-\frak{F})\wedge \beta
\right) _{\mathrm{hor}}.  \label{ASform}
\end{equation}
From (\ref{ev}) and (\ref{expequi}) we have 
\begin{equation*}
C_{f,\beta }^{\#}(X)=\int_{M}\mathrm{ev}^{\ast }(c_{f}(\mathbb{F}^{\mathrm{%
Gau}^{0}P}(X))\wedge p^{\ast }\beta )=\int_{M}c_{f}(\mathrm{ev}^{\ast }(%
\mathbb{F})-X)\wedge \beta .
\end{equation*}
So $\mathrm{ChW}_{\mathcal{A}}(C_{f,\beta }^{\#})$ is also represented by
the form (\ref{ASform}).
\end{proof}

\begin{remark}
\emph{The construction of the classes }$\mu _{f}([\beta ])$ \emph{appears in 
\cite{AS} in order to compute the Chern character of the index of families
of Dirac operators and to apply them to the study of anomalies in gauge
theories. These classes also appear in other constructions in gauge
theories, like the definition of Donaldson invariants (see \cite{Donaldson}%
), Topological Quantum Field Theory (\cite{TFT,BS1}), etc.}

\emph{It is remarkable that we obtain these classes only by studying }$%
\mathrm{Gau}P$\emph{-invariant forms on }$C(P)$\emph{\ and its equivariant
extensions.}
\end{remark}

\subsection{Moment maps\label{subsec3.2}}

\begin{example}
\label{simpectica}\emph{Let} $M$\emph{\ be a surface, and }$G=U(k)$. \emph{If%
} $f(X)=\tfrac{1}{8\pi ^{2}}\mathrm{tr}(X^{2})$\emph{, }$X\in \frak{g}$\emph{%
,} \emph{then the corresponding characteristic class on }$M$\emph{\ vanishes
by dimensional reasons, but the characteristic form }$c_{f}(\mathbb{F})\in
\Omega ^{4}(C(P))\ $\emph{does not. From our constructions, this form
defines a closed and }$\mathrm{Gau}P$\emph{-invariant }$2$\emph{-form on} $%
\mathcal{A}$, $C_{f}=\digamma \lbrack c_{f}(\mathbb{F})]$\emph{. By
Proposition \ref{expromega} and formula (\ref{expequi}) for} $a,b\in \Omega
^{1}(M,\mathrm{ad}P)$\emph{, and} $X\in \mathrm{gau}P=\Omega ^{0}(M,\mathrm{%
ad}P)$\emph{,} \emph{we have } 
\begin{eqnarray*}
c_{f}(\mathbb{F}) &=&\tfrac{1}{8\pi ^{2}}\mathrm{tr}(\mathbb{F}\wedge 
\mathbb{F}), \\
c_{f}(\mathbb{F}^{\mathrm{Gau}P})(X) &=&\tfrac{1}{8\pi ^{2}}\mathrm{tr}(%
\mathbb{F}\wedge \mathbb{F})-\tfrac{1}{4\pi ^{2}}\mathrm{tr}(p^{\ast }X\cdot 
\mathbb{F})+\tfrac{1}{8\pi ^{2}}\mathrm{tr}((p^{\ast }X)^{2}), \\
(C_{f})_{A}(a,b) &=&\tfrac{1}{4\pi ^{2}}\int_{M}\mathrm{tr}(a\wedge b), \\
(C_{f}^{\#}(X))_{A}(a,b) &=&\tfrac{1}{4\pi ^{2}}\int_{M}\mathrm{tr}(a\wedge
b)-\tfrac{1}{4\pi ^{2}}\int_{M}\mathrm{tr}(X\cdot F_{A}).
\end{eqnarray*}
\emph{Hence,} $C_{f}$ \emph{coincides with the natural symplectic structure
on} $\mathcal{A}$ \emph{defined in \cite{AB1}. Moreover, in the case of a} $%
2 $\emph{-form, it is equivalent to give an equivariant extension of the
form and a moment map (}e.g. \emph{see \cite{AB2}). Hence} $C_{f}^{\#}$ 
\emph{defines a canonical moment map} $m$ \emph{for this symplectic
structure, given by,} 
\begin{eqnarray*}
m\colon \mathcal{A} &\rightarrow &(\mathrm{gau}P)^{\ast }, \\
m_{A}(X) &=&-\tfrac{1}{4\pi ^{2}}\int_{M}\mathrm{tr}(X\cdot F_{A}).
\end{eqnarray*}
\emph{Under the pairing } 
\begin{eqnarray}
\Omega ^{2}(M,\mathrm{ad}P)\times \Omega ^{0}(M,\mathrm{ad}P) &\rightarrow &%
\mathbb{R}  \label{pairing} \\
(\eta ,X) &\mapsto &\left\langle \eta ,X\right\rangle =-\tfrac{1}{4\pi ^{2}}%
\int_{M}\mathrm{tr}(X\cdot \eta ).  \notag
\end{eqnarray}
\emph{this moment map corresponds to the curvature }$F_{A}$\emph{, and it
thus coincides with that defined in \cite{AB1}.}

\emph{Also, we have} $m^{-1}(0)=\{A\in \mathcal{A}:$ $F_{A}=0\}$\emph{, and
by symplectic reduction we obtain the moduli space of flat connections, and
our form gives rise to the symplectic structure on this space.}
\end{example}

More generally, let $(M,\sigma )$ be a symplectic $2n$-manifold. Then the
form 
\begin{equation*}
\tfrac{1}{(n-1)!}c_{f}(\mathbb{F})\wedge \sigma ^{n-1}\in \Omega
^{2n+2}(C(P))
\end{equation*}
defines a symplectic structure on $\mathcal{A}$, and the equivariant
extension provides a moment map for it, which, in particular, coincides with
that obtained in \cite[Prop 6.5.8]{Donaldson} and \cite[sec. 3]{Leung}.

\subsection{Chern-Simons terms}

Suppose that $M$\ has dimension $2k-1$ and $f\in \mathcal{I}_{k}^{G}$. Then $%
c_{f}(\mathbb{F})\in \Omega ^{2k}(C(P))$ defines a first order locally
variational operator (see \cite{FM}). Let $h\colon \Omega ^{\bullet
}\rightarrow \Omega ^{\bullet }(J^{1}C(P))$ denote the horizontalization
operator. As we have $c_{f}(\mathbb{F})=d\eta _{f}(A_{0})$, the form $h(\eta
_{f}(A_{0}))\in \Omega ^{2k-1}(J^{1}(C(P)))$, is a Lagrangian for this
operator, and hence this operator is globally variational.

We know that $c_{f}(\mathbb{F})$ is $\mathrm{Gau}P$-invariant, but $\eta
_{f}(A_{0})$ is not invariant, because it depends on the connection $A_{0}$.
However, by virtue of (\ref{descenso}) for every $X\in \mathrm{gau}P$, we
have 
\begin{eqnarray*}
L_{X_{C(P)}}\eta _{f}(A_{0}) &=&i_{X_{C(P)}}d\eta
_{_{f}}(A_{0})+di_{X_{C(P)}}\eta _{_{f}}(A_{0}) \\
&=&i_{X_{C(P)}}c_{f}(\mathbb{F})+di_{X_{C(P)}}\eta _{_{f}}(A_{0}) \\
&=&d\left( c_{f}^{1}(\mathbb{F},X)+i_{X_{C(P)}}\eta _{_{f}}(A_{0})\right) ,
\end{eqnarray*}
and hence $L_{X_{C(P)}}\eta _{f}(A_{0})$ si exact. As it is shown in \cite
{GMS} this condition leads to a Noether conservation law. In fact, the
conserved current is $\frak{J}(X)=h(c_{f}^{1}(\mathbb{F},X))$, because by
the results in \cite{FM} $A$ is a extremal connection if and only if $\sigma
_{A}^{\ast }(i_{Y}c_{f}(\mathbb{F}))=0$, $\forall Y\in \frak{X}(C(P))$, and
in this case, for any $X\in \mathrm{gau}P$, we have 
\begin{equation*}
d\sigma _{A}^{\ast }(c_{f}^{1}(\mathbb{F},X))=\sigma _{A}^{\ast }d(c_{f}^{1}(%
\mathbb{F},X))=\sigma _{A}^{\ast }(i_{X_{C(P)}}c_{f}(\mathbb{F}))=0.
\end{equation*}

More generally, if $f\in \mathcal{I}_{k}^{G}$,the form $\beta \in \Omega
^{r}(M)$ is closed, and $\dim (M)=2k+r-1$, the form $c_{f}(\mathbb{F})\wedge
p^{\ast }\beta $ defines a first order globally variational operator with
lagrangian density $\lambda =h\left( \eta _{f}(A_{0})\right) \wedge p^{\ast
}\beta $ and with conserved current $\frak{J}(X)=h(c_{f}^{1}(\mathbb{F}%
,X))\wedge p^{\ast }\beta $.

The form $c_{f}(\mathbb{F})$ defines a closed and $\mathrm{Gau}P$-invariant $%
1$-form $\digamma \lbrack c_{f}(\mathbb{F})]$ on the space of connections $%
\mathcal{A}$. This form is also horizontal, because for every $X\in \mathrm{%
gau}P$ we have 
\begin{equation*}
i_{X_{\mathcal{A}}}\digamma \lbrack c_{f}(\mathbb{F})]=\digamma \lbrack
i_{X_{C(P)}}c_{f}(\mathbb{F})]=\digamma \left[ dc_{f}^{1}(X,\mathbb{F})%
\right] =0.
\end{equation*}
So, $\digamma \lbrack c_{f}(\mathbb{F})]$ projects to a closed $1$-form $%
\alpha _{f}$ on the space $\mathcal{A}/\mathrm{Gau}^{0}P$. We have 
\begin{equation*}
\digamma \lbrack c_{f}(\mathbb{F})]=\digamma \lbrack d\eta
_{f}(A_{0})]=d\digamma \lbrack \eta _{f}(A_{0})].
\end{equation*}
Hence $\digamma \lbrack c_{f}(\mathbb{F})]$ is the exterior differential of
the function $\digamma \lbrack \eta _{f}(A_{0})]\in \Omega ^{0}(\mathcal{A})$%
. It is easy to see that the $1$-form $\alpha _{f}\in \Omega ^{1}(\mathcal{A}%
/\mathrm{Gau}^{0}P)$ is exact if and only if the function $\digamma \lbrack
\eta _{f}(A_{0})]$ is $\mathrm{Gau}^{0}P$-invariant. We have 
\begin{equation*}
L_{X_{\mathcal{A}}}\digamma \lbrack \eta _{f}(A_{0})]=\digamma \lbrack
L_{X_{C}}\eta _{f}(A_{0})]=\digamma \left[ d\left( c_{f}^{1}(\mathbb{F}%
,X)+i_{X_{C}}\eta _{_{f}}(A_{0})\right) \right] =0,
\end{equation*}
and so this function is invariant under the action of the connected
component with the identity in $\mathrm{Gau}P$. But in general it is not
invariant under the action of the full group $\mathrm{Gau}^{0}P$ (as it is
shown in the following example),\ and in this case $\alpha _{f}$ defines a
non-trivial cohomology class on $\mathcal{A}/\mathrm{Gau}^{0}P$.

\begin{example}
\emph{Suppose that }$G=SU(2)$\emph{,} $f=\frac{1}{4\pi ^{2}}\mathrm{det}$ 
\emph{is the polynomial corresponding to the second Chern class and} $M$%
\emph{\ is a} $3$\emph{-manifold. Then the bundle }$P$ \emph{is trivial }$%
P=M\times SU(2)$\emph{,} $\mathcal{A}\cong \Omega ^{1}(M,\frak{g})$ \emph{and%
} $\mathrm{Gau}P=C^{\infty }(M,SU(2))$\emph{. If }$A_{0}$ \emph{is the
connection corresponding to the product decomposition, then} $\eta
_{_{f}}(A_{0})$\emph{\ is the classical Chern-Simons Lagrangian and for any} 
$A\in \mathcal{A}$ \emph{we have} 
\begin{equation*}
\digamma \lbrack \eta _{f}(A_{0})]_{A}=\int_{M}\sigma _{A}^{\ast }\left(
\eta _{f}(A_{0})\right) =-\tfrac{1}{8\pi ^{2}}\int_{M}\left( A\wedge dA+%
\tfrac{2}{3}A\wedge A\wedge A\right)
\end{equation*}
\emph{If} $\varphi \colon M\rightarrow SU(2)$ \emph{is a gauge
transformation, it is a classical result (see \cite{TFT}) that} 
\begin{equation*}
\digamma \lbrack \eta _{f}(A_{0})]_{\varphi \cdot A}=\digamma \lbrack \eta
_{f}(A_{0})]_{A}-S(\varphi ),
\end{equation*}
\emph{where} $S(\varphi )$\emph{\ is the winding number of the map} $\varphi 
$\emph{. As} $SU(2)$\emph{\ is connected, every gauge transformation is
homotopic to an element of }$\mathrm{Gau}^{0}P$\emph{. Hence there are
elements }$\varphi \in \mathrm{Gau}^{0}P$ \emph{with} $S(\varphi )\neq 0$%
\emph{, and} $\digamma \lbrack \eta _{f}(A_{0})]$\emph{\ is not }$\mathrm{Gau%
}^{0}P$\emph{-invariant.}
\end{example}

\section{Concluding remarks}

\begin{enumerate}
\item  The Berline-Vergne definition of equivariant characteristic classes
supposes that a $\mathcal{G}$-invariant connection is given. Note however
that, independently of the existence of $\mathcal{G}$-invariant
connections,\ the $\mathcal{G}$-equivariant characteristic forms always
exist on $C(P)$ (since the canonical connection is $\mathcal{G}$-invariant),
and the existence of $\mathcal{G}$-invariant connections is needed only in
order to obtain $\mathcal{G}$-equivariant classes on $M$. We hope that our
construction could be useful in the study of equivariant characteristic
classes for non-compact Lie groups, where the existence of invariant
connections is not guaranteed in general, and the analysis is much more
involved; \emph{e.g.},\emph{\ }see \cite{GET}.

Moreover, in Section \ref{sec3} we have used $\mathrm{Gau}P$-equivariant
characteristic forms. From the classical point of view of equivariant
characteristic classes this procedure is meaningless as this group acts
trivially on $M$ and also there are no $\mathrm{Gau}P$-invariant connections.

\item  The usefulness of the map $\digamma $ lies in the fact that it
provides a general procedure to obtain results about (equivariant)
differential forms and cohomology classes on the infinite dimensional
manifold $\Gamma (E)$ by working on a finite dimensional jet bundle. Note
that, as in this paper we only consider forms on the $0$-jet bundle (that
is, on $E$), it could be think that the consideration of jet bundles is
unnecessary; but, for example in \cite{equimet}, we study the analogous
results in the case of Riemannian metrics, and in this case we need to work
with forms in the first jet bundle. In fact, there is a close relation
between the map $\digamma $ and the variational bicomplex, that we will
analyze in a forthcoming paper.
\end{enumerate}

\end{document}